\documentclass{sig-alternate-2013}
\usepackage{url}
\usepackage{balance}
\usepackage{booktabs}
\usepackage{hyphsubst}
\usepackage{graphicx}
\usepackage{amsmath}
\graphicspath{{images/}}
\usepackage{times}
\usepackage{hhline}
\usepackage[caption = false]{subfig}

\permission{Permission to make digital or hard copies of all or part of this work for personal or classroom use is granted without fee provided that copies are not made or distributed for profit or commercial advantage and that copies bear this notice and the full citation on the first page. Copyrights for components of this work owned by others than ACM must be honored. Abstracting with credit is permitted. To copy otherwise, or republish, to post on servers or to redistribute to lists, requires prior specific permission and/or a fee. Request permissions from permissions@acm.org.}
\conferenceinfo{HT'14,}{Santiago, Chile, September 01-04, 2014.}
\copyrightetc{Copyright 2014 ACM \the\acmcopyr}
\crdata{xxx-x-xxxx-xxxx-x/xx/xx\ ...\$xx.xx}

\begin{document}

\title{SocRecM: A Scalable Social Recommender Engine for Online Marketplaces}

\numberofauthors{4} 
\author{
\alignauthor
Emanuel Lacic\\
       \affaddr{Graz University of Technology}\\
       \affaddr{Graz, Austria}\\
       \email{elacic@know-center.at}
\alignauthor
Dominik Kowald\\
\affaddr{Know-Center}\\
       \affaddr{Graz, Austria}\\
       \email{dkowald@know-center.at}
\and
\alignauthor 
Christoph Trattner\\
\affaddr{Know-Center}\\
       \affaddr{Graz, Austria}\\
       \email{ctrattner@know-center.at}
}

\maketitle

\begin{abstract}
In this paper, we present work-in-progress on \textit{SocRecM}, a novel social recommendation framework for online marketplaces. We demonstrate that \textit{SocRecM} is not only easy to integrate with existing Web technologies through a RESTful, scalable and easy-to-extend service-based architecture but also reveal the extent to which various social features and recommendation approaches are useful in an online social marketplace environment.
\end{abstract}


\section{Introduction} \label{sec:introduction}

Recommender systems aim at helping users find relevant information
in an overloaded information space. Although various recommender frameworks are available nowadays, there is still a lack of frameworks that address important aspects in recommender systems research such as: easy integration in an existing infrastructure, scalability, hybridization and social data integration.
To tackle these issues, we implemented \textit{SocRecM}, a scalable and easy-to-integrate online social recommender framework whose purpose is not only to decrease the workload of developers via an easy-to-use framework but also to provide recommendation algorithms that utilize social data obtained from various data sources (e.g., Facebook, Twitter, etc.).


\begin{figure}[ht!]
  \centering
    \includegraphics[width=.45\textwidth]{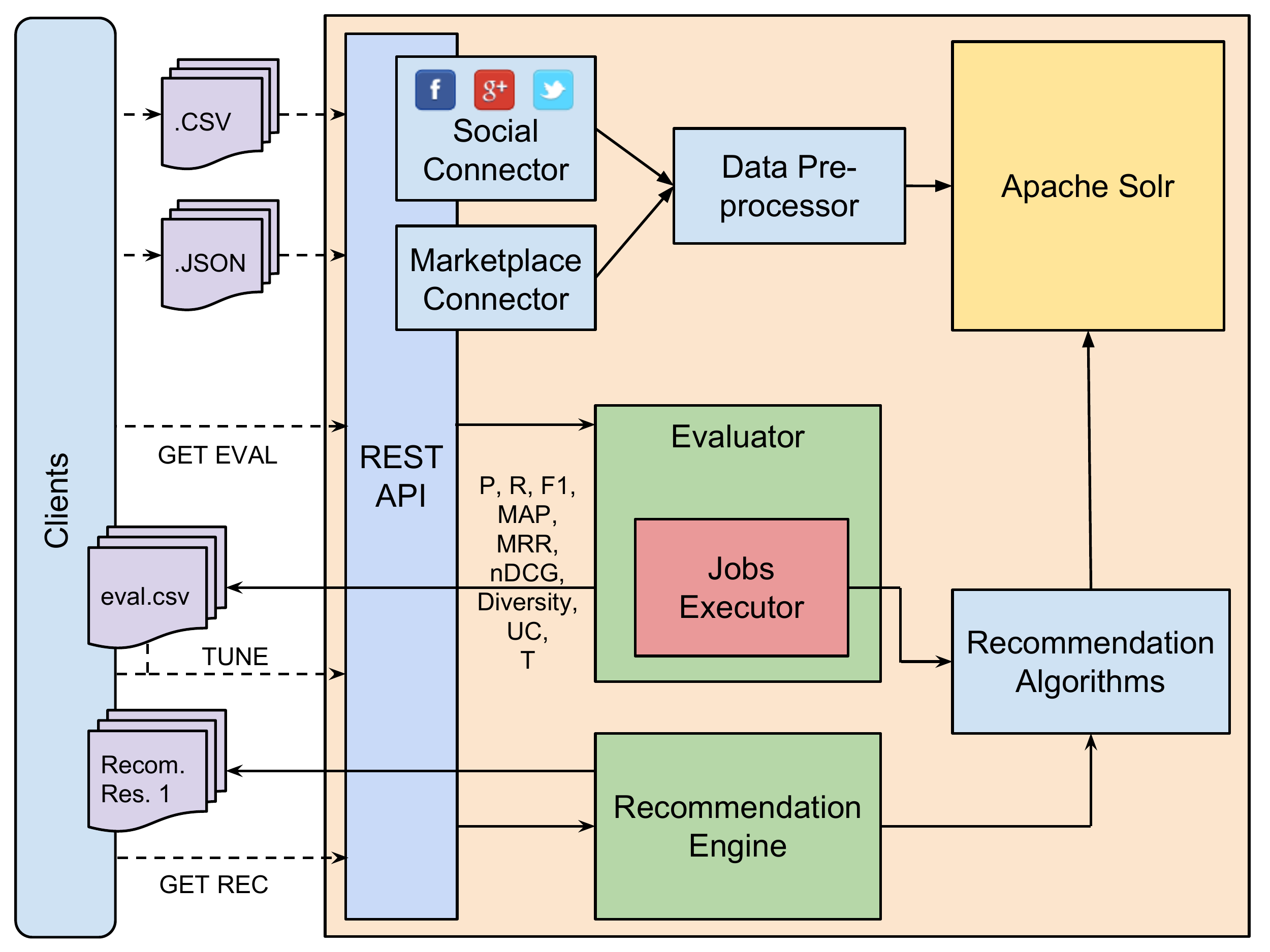}
      \caption{\textit{SocRecM} system architecture.}
    \label{fig:architecture}
\end{figure}

\section{System Overview}
\label{sec:approach}
The first prototype of \textit{SocRecM} was implemented in Java and can be found online as open-source software \footnote{https://github.com/lacic/solr-resource-recommender}. 
Figure \ref{fig:architecture} shows the system architecture of \textit{SocRecM}. As featured, the engine can easily be integrated into a RESTful API for communicating with client applications. The \textit{SocRecM} API provides methods for uploading marketplace and social data into the engine and for querying resource recommendations and benchmarking results. The marketplace and social data is gathered via respective connectors and pre-processed to be indexed by Apache Solr, which in turn offers powerful search and content analyzing functionalities (e.g., facets or MoreLikeThis queries) that are used by the recommendation algorithm implementations. Currently, \textit{SocRecM} contains four types of algorithms to recommend resources (in our case products) to users, including MostPopular (MP), Collaborative Filtering (CF) \cite{schafer2007collaborative}, Content-based (C) \cite{pazzani2007content} and Hybrid Recommendations (CCF) \cite{burke2002hybrid} (see also \cite{lacictowards}). These algorithms are calculated based on either marketplace features, such as purchases (CF$_p$), title (C$_t$) and  description (C$_{d}$) or social features, such as likes (CF$_{l}$), comments (CF$_{c}$), interactions(CF$_{in}$), social stream content data (C$_{st}$), groups (CF$_{g}$) and interests (CF$_{i}$). Additionally, the algorithms and data features are incorporated into hybrid algorithms (CCF$_m$ for marketplace and CCF$_s$ for social features).

The recommendation algorithms are invoked by the evaluator component and its jobs executor in order to evaluate them with respect to various IR metrics (e.g., Recall (R), Precision (P), nDCG, User Coverage (UC), etc.). The evaluation results are further used in \textit{SocRecM} to tune the parameters of the algorithms, especially in the case of the hybrid approaches. Furthermore, the algorithms are called by the recommendation engine in order to forward the recommended resources back to the client.

\begin{figure}[t!]
  \centering
		 \subfloat[Rec. quality (experiment 1)]{ 
				\includegraphics[width=0.24\textwidth]{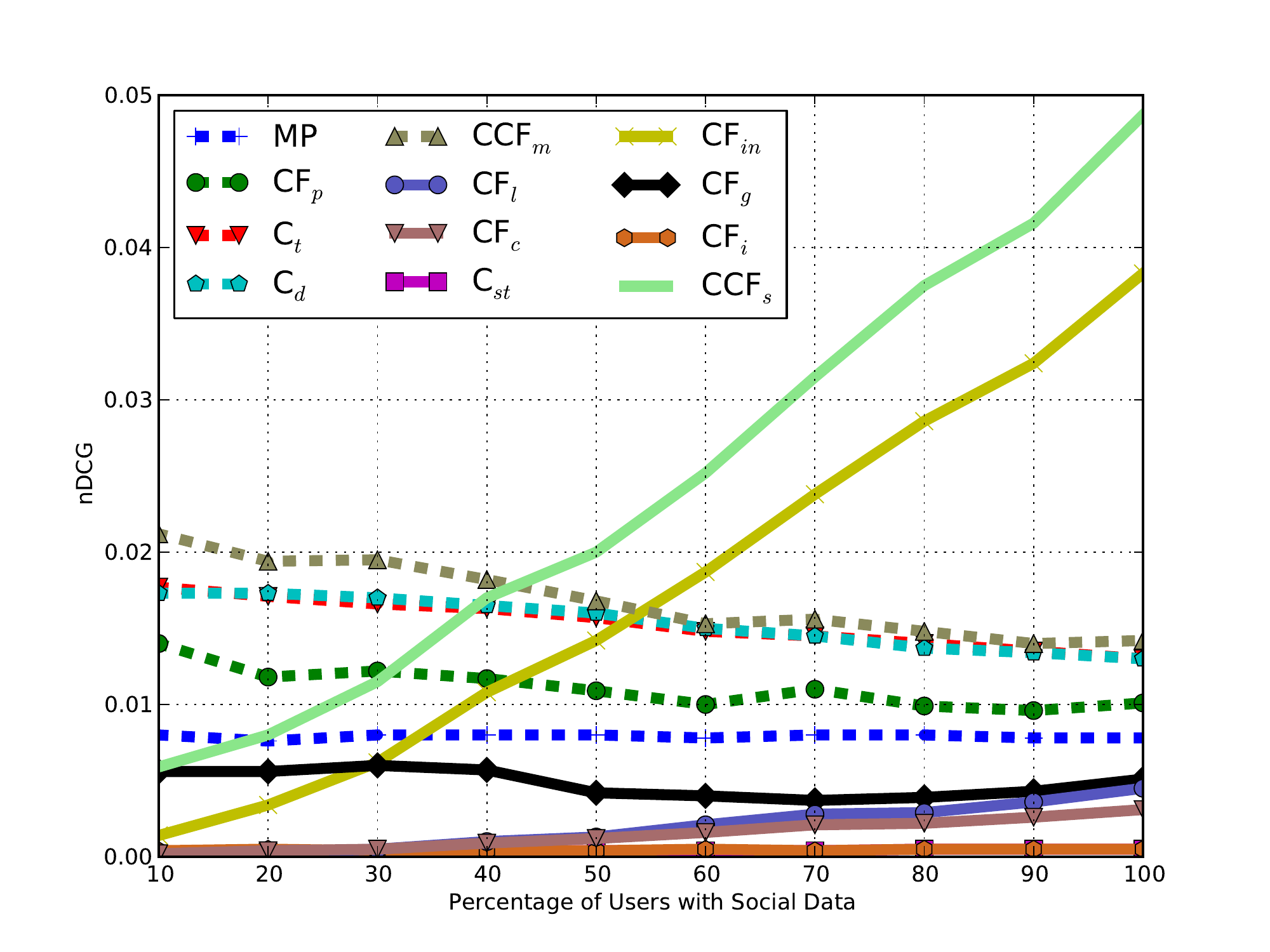}%
		 } 
		 \subfloat[UC (experiment 1)]{ 
				\includegraphics[width=0.24\textwidth]{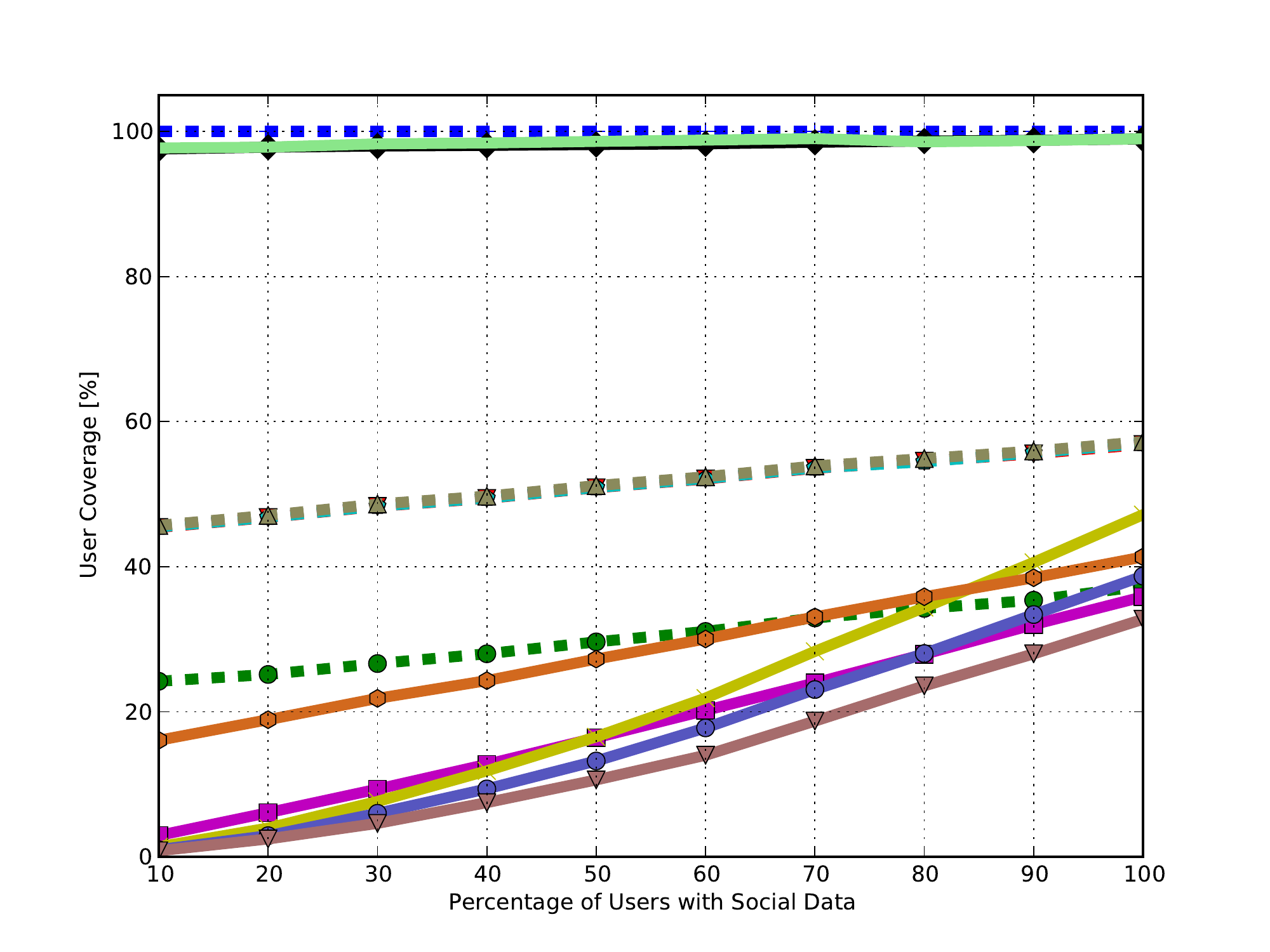}%
		 }  \\
		 \subfloat[Rec. quality (experiment 2)]{ 
				\includegraphics[width=0.24\textwidth]{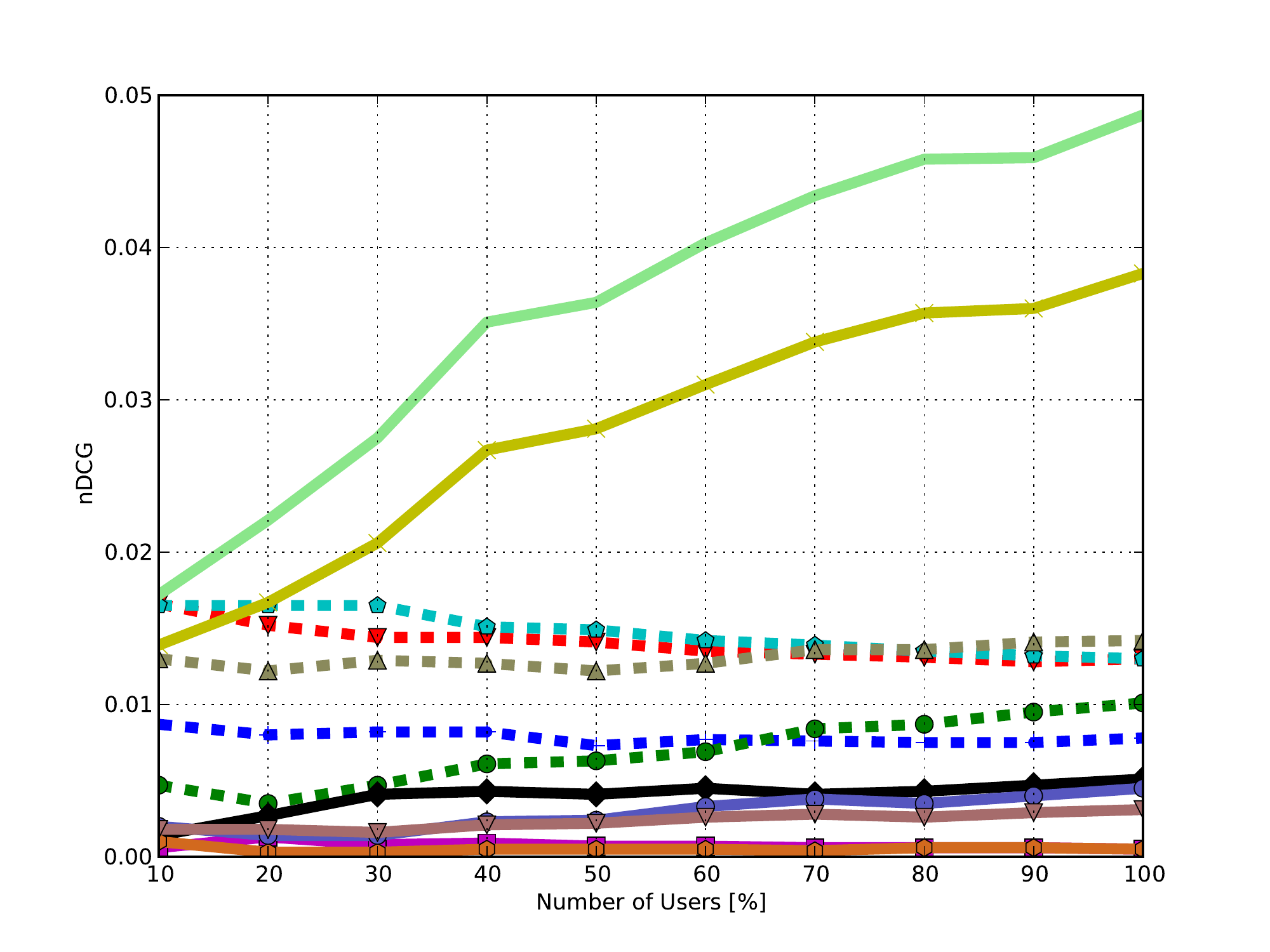}%
		 } 
		 \subfloat[UC (experiment 2)]{ 
				\includegraphics[width=0.24\textwidth]{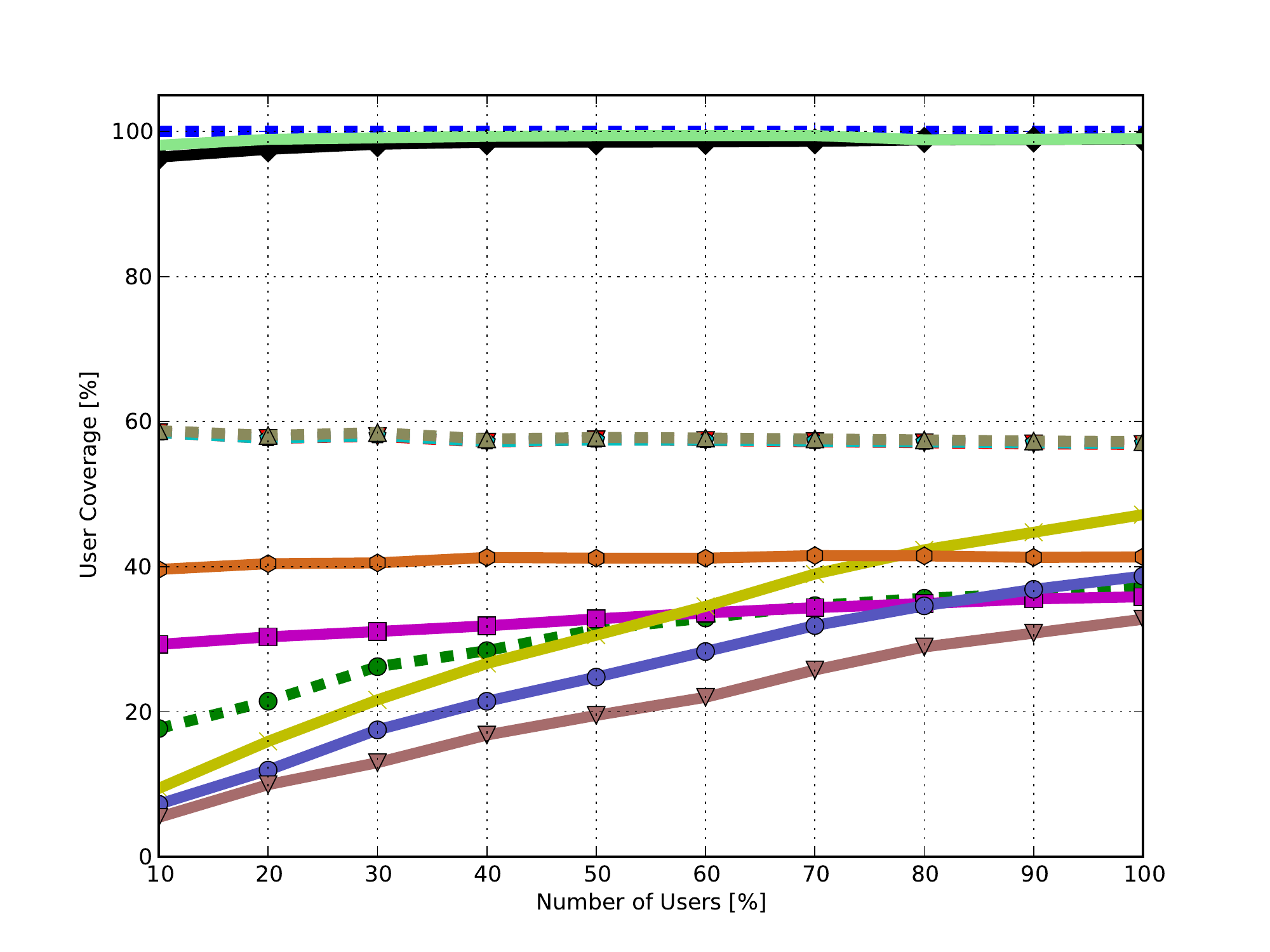}%
		 } 
     \caption{Results of the two experiments in terms of recommendation quality (nDCG) and User Coverage (UC).}
    \label{fig:results}
\end{figure}

%

\section{Results}
\label{sec:eval}
To demonstrate the effectiveness of our approach under different scenarios we conducted two ``virtual'' experiments with a social dataset gathered from the virtual world of SecondLife, which provides both: detailed marketplace purchase data and social data collected from users in the virtual world. The dataset contains 126,356 users, 122,360 products, 265,274 purchases, 1,839,783 social interactions, 510,145 social stream contents, 260,137 groups and 88,371 interests and was split into a training and test set using the method proposed in \cite{lacictowards}. The following experiments were conducted using an IBM System x3550 server with two 2.0 GHz six-core Intel Xeon E5-2620 processors, a 1TB ServeRAID M1115 SCSI Disk and 128 GB of RAM using one Apache Solr 4.3.1 instance in the back-end. 

In the first experiment we show how the recommendation quality in terms of the nDCG and User Coverage (UC) metrics \cite{Jarvelin02} can be improved if the users' social data is provided in addition to the marketplace data. To simulate this, we extracted all users from our dataset that have both the marketplace and social data (10,996 users). Out of this subset, we randomly selected 10\% of users and replaced them with users that only provided marketplace data. We continued until 100\% of the data consisted of users with only marketplace data in their profiles. As demonstrated in plots (a) and (b) of Figure \ref{fig:results}, the recommendation approaches based on marketplace data alone are fairly constant regardless of how many users have social data in their profiles. This is expected since the number of users neither increased nor decreased throughout the experiment. However, it is apparent that the recommenders based on the users' interactions perform best (CCF$_s$ and CF$_{in}$) but are also significantly depended on the number of social profiles in the dataset. An interesting finding in this context is that the content-based approach based on the users' social stream contents (CF$_{st}$) performed as poorly as the recommender based on the users' interests (CF$_{i}$).

In the second experiment we simulated a cold-start scenario for a new social marketplace system, under which we assume that all new users provide social data as it was the case with the start of Spotify, to determine if and how the recommendation quality would be affected by an increasing number of social users. In order to conduct this experiment, we again extracted all users from the dataset with both the marketplace and the social data (10,996) and eliminated the rest. Using these data, we created 10 different sets with an increasing number of users and evaluated the recommendation approaches on them, as demonstrated in plots (c) and (d) in Figure \ref{fig:results}. As the number of users with marketplace and social data increased, the approaches based on the social data (CCF$_s$ and CF$_{in}$) delivered a much higher prediction quality than those that only used the marketplace data (e.g., CF$_p$). Moreover, the user coverage of the hybrid approaches based on social data (CCF$_s$) is much higher than the one based on the marketplace data (CCF$_m$). Comparing CF$_p$ with CF$_{in}$, shows that although both algorithms get a higher user coverage with more users, the social features ultimately provide better results.

\begin{figure}
\vspace{5mm}
  \centering
    \includegraphics[width=0.45\textwidth]{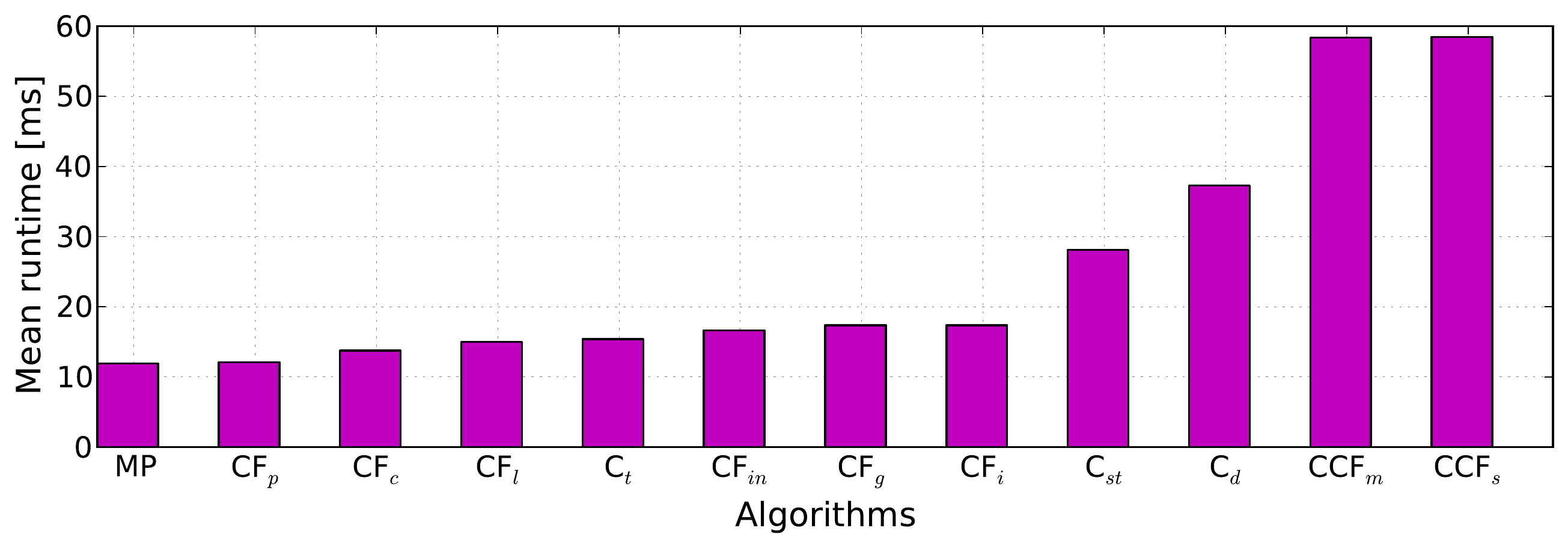}
      \caption{Mean runtime of the algorithms in \textit{SocRecM}.}
    \label{fig:runtime}
\end{figure}

In addition to the recommendation quality, we compared the mean runtime (i.e., the time needed to calculate recommendations for a user) of the recommendation approaches shown in Figure \ref{fig:runtime}. In general these results demonstrate that \textit{SocRecM} is capable of providing near real-time recommendations for users since the maximum mean test time was only 58 milliseconds for the hybrid approaches CCF$_m$ and CCF$_s$.

\section{Future Work} \label{sec:con}
In the future, we plan to further extend our framework using different approaches to make sense of social data provided by users in a social marketplace environment. For example, we would like to implement a topic modeling approach based on LDA to combine comments with purchase data or to derive topics from the user's social stream to calculate similarities between users. Furthermore, we are interested in extending our framework to generate recommendations based on the user's geo-location data. Last but not least, we are interested in developing novel hybridization approaches for diverse social data sources to further increase the predictive power of our recommender framework.

\textbf{Acknowledgments:}
This work is supported by Know-Center and the EU-funded project Learning Layers (Grant Nr. 318209).

\balance
\bibliographystyle{abbrv}
\small
\bibliography{ht14resource_poster}

\end{document}